\documentclass[onecolumn,showpacs,preprintnumbers,amsmath,amssymb]{revtex4}

\usepackage{epsf}
\usepackage{graphicx}  
\usepackage{dcolumn}   
\usepackage{bm}        
\usepackage{subfigure}

\def\etal{et al.~}

\begin{document}

\title{Hubble parameter data constraints on dark energy}

\author{Yun Chen}
 \email{chenyun@mail.bnu.edu.cn}
\affiliation{Department of Astronomy, Beijing Normal University,
Beijing 100875, China}
 \affiliation{Department of Physics, Kansas State University, 116
Cardwell Hall, Manhattan, KS 66506, USA}

\author{Bharat Ratra}
\email{ratra@phys.ksu.edu}
 \affiliation{Department of Physics, Kansas State University, 116
Cardwell Hall, Manhattan, KS 66506, USA}

\date{ \today}


\begin{abstract}

 We use Hubble parameter versus redshift data from Stern \etal
 \cite{Stern2010} and Gazta\~{n}aga \etal \cite{gaztanaga09} to place constraints on
model parameters of constant and time-evolving dark energy
cosmological models. These constraints are consistent with (through
not as restrictive as) those derived from supernova Type Ia
magnitude-redshift data. However, they are more restrictive than
those derived from galaxy cluster angular diameter distance, and
comparable with those from gamma-ray burst and lookback time data. A
joint analysis of the Hubble parameter data with more restrictive
baryon acoustic oscillation peak length scale and supernova Type Ia
apparent magnitude data favors a spatially-flat cosmological model
currently dominated by a time-independent cosmological constant but
does not exclude time-varying dark energy.

\end{abstract}


\maketitle
\section{INTRODUCTION}
\label{intro}

It is well-established that the Universe is currently undergoing
accelerated cosmological expansion. Observational evidence for the
accelerated expansion comes from supernova Type Ia (SNIa) apparent
magnitude measurements as a function of redshift
\cite{Amanullah2010, shafieloo09}, cosmic microwave background (CMB)
anisotropy data \cite{ratra99} combined with low estimates of the
cosmological mass density \cite{chen03b}, and baryon acoustic
oscillation (BAO) peak length scale estimates \cite{gaztanaga09,
Percival2010, samushia09}.

The underlying mechanism responsible for this accelerated expansion
is not yet well characterized. The ``standard'' general relativistic
model of cosmology has an energy budget that is currently dominated
by far by dark energy, a negative-pressure substance that powers the
accelerated expansion. (Another possibility is that the above
observations are a manifestation of the breakdown of general
relativity on large cosmological length scales. In this paper we
assume that general relativity provides an adequate description of
gravitation on cosmological length scales.) Dark energy can vary
weakly in space and evolve slowly in time, though current data are
consistent with it being a cosmological constant. For recent reviews
see \cite{ratra08}.

There are many dark energy models under discussion (for recent
discussions see \cite{hirano11}, and references therein). The
current ``standard'' model is the $\Lambda$CDM model
\cite{peebles84} where the accelerated cosmological expansion is
powered by Einstein's cosmological constant, $\Lambda$, a spatially
homogeneous fluid with equation of state parameter $\omega_\Lambda =
p_\Lambda/\rho_\Lambda = -1$ (where $p_\Lambda$ and $\rho_\Lambda$
are the fluid pressure and energy density). In this model the
cosmological energy budget is dominated by far by $\rho_\Lambda$,
with cold dark matter (CDM) being the second largest contributor.
The $\Lambda$CDM model provides a reasonable fit to most
observational constraints, although the ``standard'' CDM structure
formation model might be in some observational trouble (see, e.g.,
\cite{Peebles2003}). In addition, the $\Lambda$CDM model raises some
puzzling conceptual questions.

If the dark energy density slowly decreased in time (rather than
remaining constant like $\rho_\Lambda$), the energy densities of
dark energy and nonrelativistic matter (CDM and baryons) would
remain comparable for a longer period of time, and so alleviate what
has become known as the $\Lambda$CDM coincidence puzzle. In
addition, a slowly decreasing effective dark energy density, based
on a more fundamental physics model that is applicable at an energy
density scale much larger than an meV, could result in the current
observed dark energy density scale of order an meV through gradual
decrease over the long lifetime of the Universe, another unexplained
feature in the context of the $\Lambda$CDM model. Thus a slowly
decreasing dark energy density could resolve some of the puzzles of
the $\Lambda$CDM model \cite{Ratra&Peebles1988}.

The XCDM parametrization is often used to describe a slowly
decreasing dark energy density. In this parametrization the dark
energy is modeled as a spatially homogenous ($X$) fluid with an
equation of state parameter $w_X = p_X/\rho_X$, where $w_X < -1/3$
is an arbitrary constant and $p_X$ and $\rho_X$ are the pressure and
energy density of the $X$-fluid. When $w_X = -1$, the XCDM
parametrization reduces to the complete and consistent $\Lambda$CDM
model. For any other value of $w_X (< -1/3)$ the XCDM
parametrization is incomplete as it cannot describe spatial
inhomogeneities (see, e.g., \cite{ratra91}). Here we study the XCDM
parametrization only in the spatially-flat cosmological case.

The $\phi$CDM model --- in which dark energy is modelled as a scalar
field $\phi$ with a gradually decreasing (in $\phi$) potential
energy density $V(\phi)$ --- is the simplest complete and consistent
model of a slowly decreasing (in time) dark energy density. Here we
focus on an inverse power-law potential energy density $V(\phi)
\propto \phi^{-\alpha}$, where $\alpha$ is a nonnegative constant
\cite{Peebles&Ratra1988, Ratra&Peebles1988}. When $\alpha = 0$ the
$\phi$CDM model reduces to the corresponding $\Lambda$CDM case. Here
we only consider the spatially-flat $\phi$CDM cosmological model.

It has been known for some time that a spatially-flat $\Lambda$CDM
model with current energy budget dominated by a constant $\Lambda$
is largely consistent with most observational constraints (see,
e.g., \cite{jassal10, allen08}). SNeIa, CMB, and BAO measurements
mentioned above indicate that we live in a spatially-flat
$\Lambda$CDM model with nonrelativistic matter contributing a little
less than 30 \% of the current cosmological energy budget, with the
remaining slightly more than 70 \% contributed by a cosmological
constant. These three sets of data carry by far the most weight when
determining constraints on models and cosmological parameters.

Future data from space missions will significantly tighten the
constraints (see, e.g., \cite{podariu01a}). However, at present, it
is important to determine independent constraints that can be
derived from other presently available data sets. While these data
are not yet as constraining as the SNeIa, CMB and BAO data, they
potentially can reassure us (if they provide constraints consistent
with those from the better known data), or if the two sets of
constraints are inconsistent this might lead to the discovery of
hidden systematic errors or rule out the cosmological model under
consideration.

Other data that have been used to constrain cosmological parameters
include galaxy cluster gas mass fraction (e.g., \cite{allen08,
Samushia&Ratra2008}), gamma-ray burst luminosity distance (e.g.,
\cite{schaefer07, Samushia&Ratra2010}), large-scale structure (e.g.,
\cite{baldi10}), strong gravitational lensing (e.g., \cite{chae04}),
and angular size (e.g., \cite{chen03b, Chen2011b, gurvits99}) data.
While the constraints from these data are less restrictive than
those derived from the SNeIa, CMB and BAO data, both types of data
result in largely compatible constraints that generally support a
currently accelerating cosmological expansion. This gives us
confidence that the broad outlines of the ``standard'' cosmological
model are now in place.

Measurements of the Hubble parameter as a function of redshift,
$H(z)$, have also been used to constrain cosmological parameters
(see \cite{Zhangetal2010} for a review). A variant of this test uses
lookback time data (see, e.g., \cite{Samushiaetal2010,
capozziello04}). Building on the work of \cite{Jimenez2002}, Simon
\etal \cite{simon05} used the differential ages of 32 passively
evolving galaxies to determine 9 $H(z)$ measurements in the redshift
range $0.09 \leq z \leq 1.75$. Cosmological constraints derived
using these data are described in \cite{Yi2007, samushia07}; more
recent references may be traced through \cite{Zhai2010}.

Stern \etal (2010, hereafter S10) \cite{Stern2010} extended the
Simon et al.\ \cite{simon05} sample to 11 measurements of $H(z)$ in
the redshift range $0.1 \leq z \leq 1.75$. These data have been used
for cosmological tests by Shafieloo \& Clarkson \cite{shafieloo10}.
It has become common to augment the S10 data with the Gazta\~{n}aga
\etal (2009, hereafter G09) \cite{gaztanaga09} estimates of $H(z)$
determined from line-of-sight BAO peak position observations. These
data, listed in Table 1, have also been used to constrain
cosmological parameters (see, e.g., \cite{Ma2011, WangandXu2010} and
references therein). There are problems with a number of these
analyses. Some of them include both the G09 data points at $z =
0.24$ and $z = 0.43$ (which we use here), as well as the G09 single
summary data point at $z = 0.34$ that is based on exactly the same
data as the two individual points. In addition, a number of these
analyses either ignore the G09 systematic errors or incorrectly
account for them. We account for the G09 statistical and systematic
errors by combining them in quadrature; the G09 data points we list
in Table 1 are identical to those used by Ma \& Zhang \cite{Ma2011}
and Zhang et al.\ \cite{Zhangetal2010}.

In this paper we use the 13 S10 and G09 $H(z)$ measurements listed
in Table 1 to constrain the $\Lambda$CDM and  $\phi$CDM models and
the XCDM parametrization. The resulting constraints are compatible
with those derived using other techniques. We also use these $H(z)$
data in combination with BAO and SNeIa measurements to jointly
constrain cosmological parameters in these models. Adding the $H(z)$
data tightens the constraints, somewhat significantly in some parts
of parameter space for some of the models we study.

Our paper is organized as follows. In Sec.\ {\ref{equations}} we
present the basic equations of the three dark energy models we
study. Constraints from the $H(z)$ data are derived in Sec.\
{\ref{HzData}}. In Sec.\ {\ref{Joint}} we determine joint
constraints on the dark energy parameters from a combination of data
sets. We summarize our main conclusions in Sec.\ {\ref{summary}}.

%
%
\section{Basic equations of the dark energy models}
\label{equations}

The Friedmann equation of the $\Lambda$CDM model with spatial
curvature can be written as
\begin{equation}
\label{eq:LCDMFriedmann} H^2(z, H_0, \textbf{p}) =
H_0^2\left[\Omega_{m0}(1+z)^3+\Omega_{\Lambda}+
(1-\Omega_{m0}-\Omega_{\Lambda})(1+z)^2\right],
\end{equation}
where $z$ is the redshift, $H(z, H_0, \textbf{p})$ is the Hubble
parameter, $H_0$ is the Hubble constant, and the model-parameter set
is $\textbf{p} = (\Omega_{m0}, \Omega_{\Lambda})$ where
$\Omega_{m0}$ is the nonrelativistic (baryonic and cold dark) matter
density parameter and $\Omega_{\Lambda}$ that of the cosmological
constant. Throughout, the subscript $0$ denotes the value of a
quantity today. In this paper, the subscripts $\Lambda$, $X$ and
$\phi$ represent the corresponding quantities of the dark energy
component in the $\Lambda$CDM, XCDM and $\phi$CDM scenarios.

In this work, for computational simplicity, spatial curvature is set
to zero in the XCDM and $\phi$CDM cases. Then the Friedmann equation
for the XCDM parametrization is
\begin{equation}
\label{eq:XCDMFriedmann} H^2(z, H_0, \textbf{p}) =
H_0^2\left[\Omega_{m0}(1+z)^3+(1-\Omega_{m0})(1+z)^{3(1+w_X)}\right],
\end{equation}
where the model-parameter set is $\textbf{p} = (\Omega_{m0}, w_X)$.

In the $\phi$CDM model, the inverse power law potential energy
density of the scalar field adopted in this paper is $V(\phi) =
\kappa m_p^2 \phi^{-\alpha}$, where $m_p$ is the Planck mass, and
$\alpha$ and $\kappa$ are non-negative constants
\cite{Peebles&Ratra1988}. In the spatially-flat case the Friedmann
equation of the $\phi$CDM model is
\begin{equation}
\label{eq:phiCDMFriedmann} H^2(z, H_0, \textbf{p}) =
\frac{8\pi}{3m_p^2}(\rho_m + \rho_{\phi}).
\end{equation}
Here $H(z) = \dot{a}/a$ is the Hubble parameter where $a(t)$ is the
cosmological scale factor and an overdot denotes a time derivative.
The energy densities of the matter and the scalar field  are
\begin{equation}
\label{eq:rhom} \rho_m = \frac{m_p^2}{6\pi}a^{-3},
\end{equation}
and
\begin{equation}
\label{eq:rhophi} \rho_{\phi} = \frac{m_p^2}{32\pi} (\dot{\phi}^2 +
\kappa m_p^2 \phi^{-\alpha}),
\end{equation}
respectively. According to  the definition of the dimensionless
density parameter, one has
\begin{equation}
\label{eq:Omegam} \Omega_m(z) = \frac{8\pi \rho_m}{3m_p^2H^2} =
\frac{\rho_m}{\rho_m + \rho_{\phi}}.
\end{equation}
The scalar field $\phi$ obeys the differential equation
\begin{equation}
\label{eq:dotphi} \ddot{\phi} + 3 \frac{\dot{a}}{a}\dot{\phi} -
\frac{\kappa \alpha}{2} m_p^2 \phi^{-(\alpha+1)} = 0.
\end{equation}
Using Eqs.\ (\ref{eq:phiCDMFriedmann}) and (\ref{eq:dotphi}), as
well as the initial conditions described in
\cite{Peebles&Ratra1988}, one can numerically compute the Hubble
parameter $H(z)$. In this case the model-parameter set is
$\textbf{p} = (\Omega_{m0}, \alpha)$.

\section{Constraints from the H(z) data}
\label{HzData}

We use the 13 $H(z)$ measurements of S10 and G09 listed in Table 1
to constrain cosmological parameters. We constrain cosmological
parameters by minimizing $\chi_{H}^2$,
\begin{equation}
\label{eq:chi2ADD} \chi_{H}^2 (H_0, \textbf{p}) =
\sum_{i=1}^{13}\frac{[H^{\rm th} (z_i; H_0, \textbf{p})-H^{\rm
obs}(z_i)]^2}{\sigma^2_{{\rm H},i}}.
\end{equation}
Here $z_i$ is the redshift at which $H(z_i)$ has been measured,
$H^{\rm th}$ is the predicted value of $H(z)$ in the cosmological
model under consideration and $H^{\rm obs}$ is the measured value.
From $\chi_{H}^2 (H_0, \textbf{p})$ we compute the likelihood
function $L(H_0, \textbf{p})$. We then treat $H_0$ as a nuisance
parameter and marginalize over it using a gaussian prior with $H_0 =
68\pm3.5$ km s$^{-1}$ Mpc$^{-1}$ \cite{chen03, Chen2011a} to get a
likelihood function $L(\textbf{p})$ that is a function of only the
cosmological parameters of interest. The best-fit parameter values
$p*$ are those that maximize this likelihood function and the 1, 2
and 3 $\sigma$ constraint contours are the set of cosmological
parameters (centered on $p*$) that enclose 68.27, 95.45 and 99.73
\%, respectively, of the probability under the likelihood function.

\begin{figure}[t]
\centering
  \includegraphics[angle=0,width=90mm]{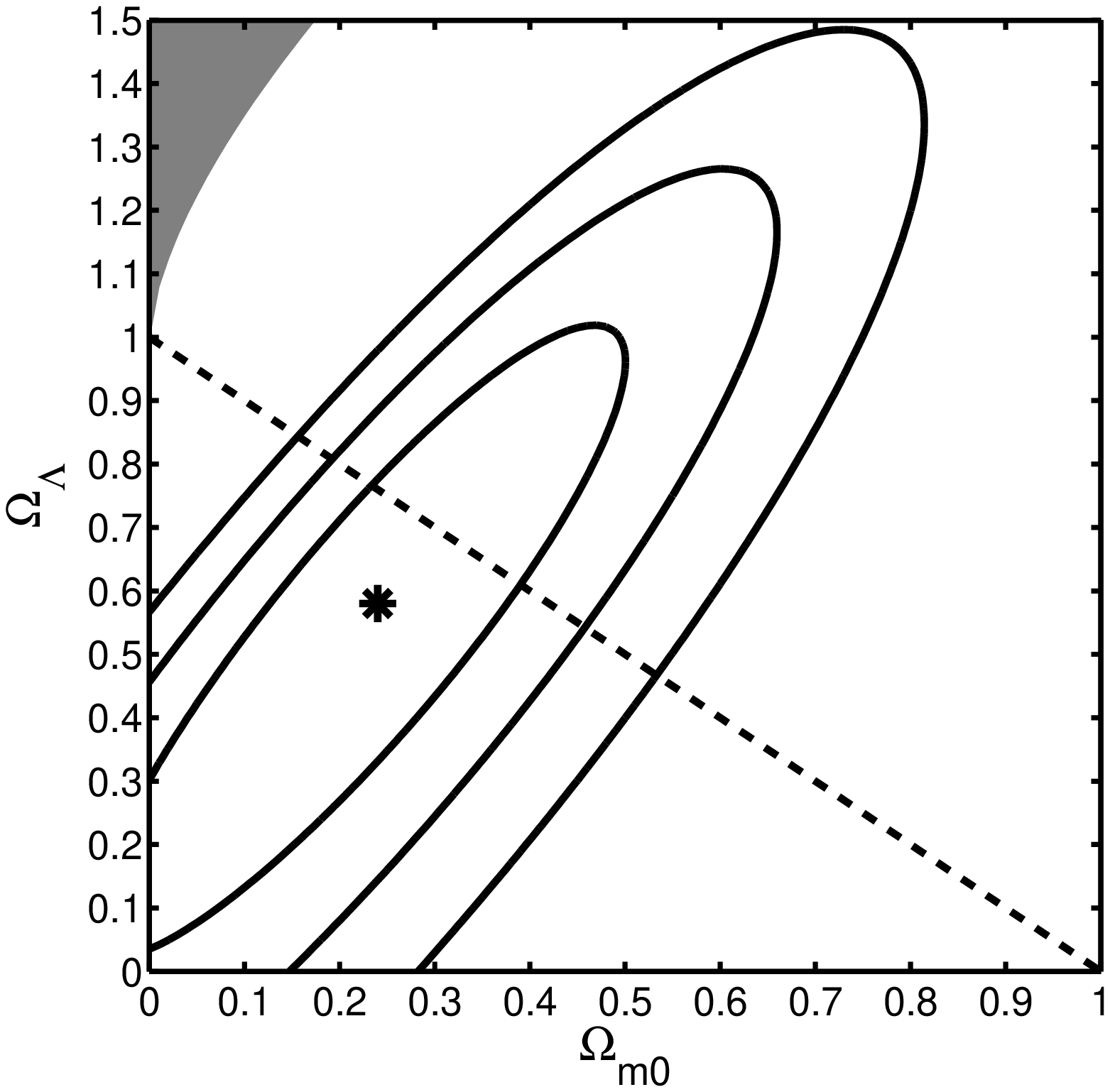}

\caption{
1, 2, and 3 $\sigma$ constraint contours for the $\Lambda$CDM model
from the $H(z)$ data. The dashed diagonal line corresponds to
spatially-flat models and the shaded area in the upper left-hand
corner is the region for which there is no big bang. The star marks
the best-fit pair $(\Omega_{m0}, \Omega_{\Lambda}) = (0.24, 0.58)$
with $\chi^2_{\rm min}=10.1$.
} \label{fig:LCDM_Hz}
\end{figure}

\begin{figure}[t]
\centering
  \includegraphics[angle=0,width=90mm]{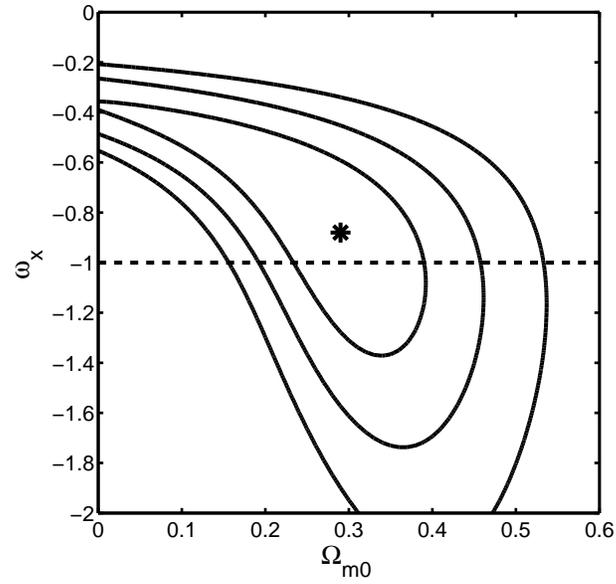}

\caption{
1, 2, and 3 $\sigma$ constraint contours for the XCDM
parametrization from the $H(z)$ data. The dashed horizontal line at
$\omega_X = -1$ corresponds to spatially-flat $\Lambda$CDM models.
The star marks the best-fit pair $(\Omega_{m0}, w_X) = (0.29,
-0.88)$ with $\chi^2_{\rm min}=10.1$.
} \label{fig:XCDM_Hz}
\end{figure}

\begin{figure}[t]
\centering
  \includegraphics[angle=0,width=90mm]{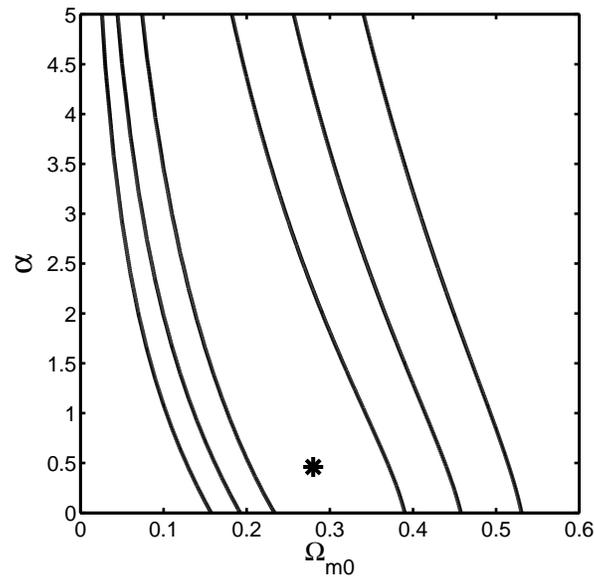}

\caption{
1, 2, and 3 $\sigma$ constraint contours for the $\phi$CDM model
from the $H(z)$ data. The horizontal axis at $\alpha = 0$
corresponds to spatially-flat $\Lambda$CDM models. The star marks
the best-fit pair $(\Omega_{m0}, \alpha) = (0.28, 0.46)$ with
$\chi^2_{\rm min}=10.1$.
} \label{fig:phiCDM_Hz}
\end{figure}

Figures \ref{fig:LCDM_Hz}--\ref{fig:phiCDM_Hz} show the constraints
from the $H(z)$ data on the three dark energy models we consider.
Not unexpectedly, these show that the $H(z)$ data fairly tightly
restrict one combination of the cosmological parameters while
leaving the ``orthogonal'' combination relatively unconstrained.
Comparing these results to those shown in Figs.\ 1--3 of
\cite{samushia07}, determined using the 9 $H(z)$ data points of
Simon et al.\ \cite{simon05}, we see that the newer S10 and G09 data
result in significantly more restrictive constraints on cosmological
parameters. These constrains are comparable with those from
gamma-ray burst (\cite{Samushia&Ratra2010}, Figs.\
1--3) and lookback time (\cite{Samushiaetal2010},
Figs.\ 1--3) data. They are more restrictive than those that follow
from galaxy cluster angular diameter distance data
(\cite{Chen2011b}, Figs.\ 1--3).

\section{Joint constraints}
\label{Joint}

Following \cite{Chen2011b}, we derive constraints on cosmological
parameters of the three models from a joint analysis of the $H(z)$
data with the BAO data \cite{Percival2010} and Union2 compliation of
557 SNeIa apparent magnitude measurements (covering a redshift range
$0.015<z <1.4$) from \cite{Amanullah2010}.

\begin{figure}[t]
\centering
  \includegraphics[angle=0,width=90mm]{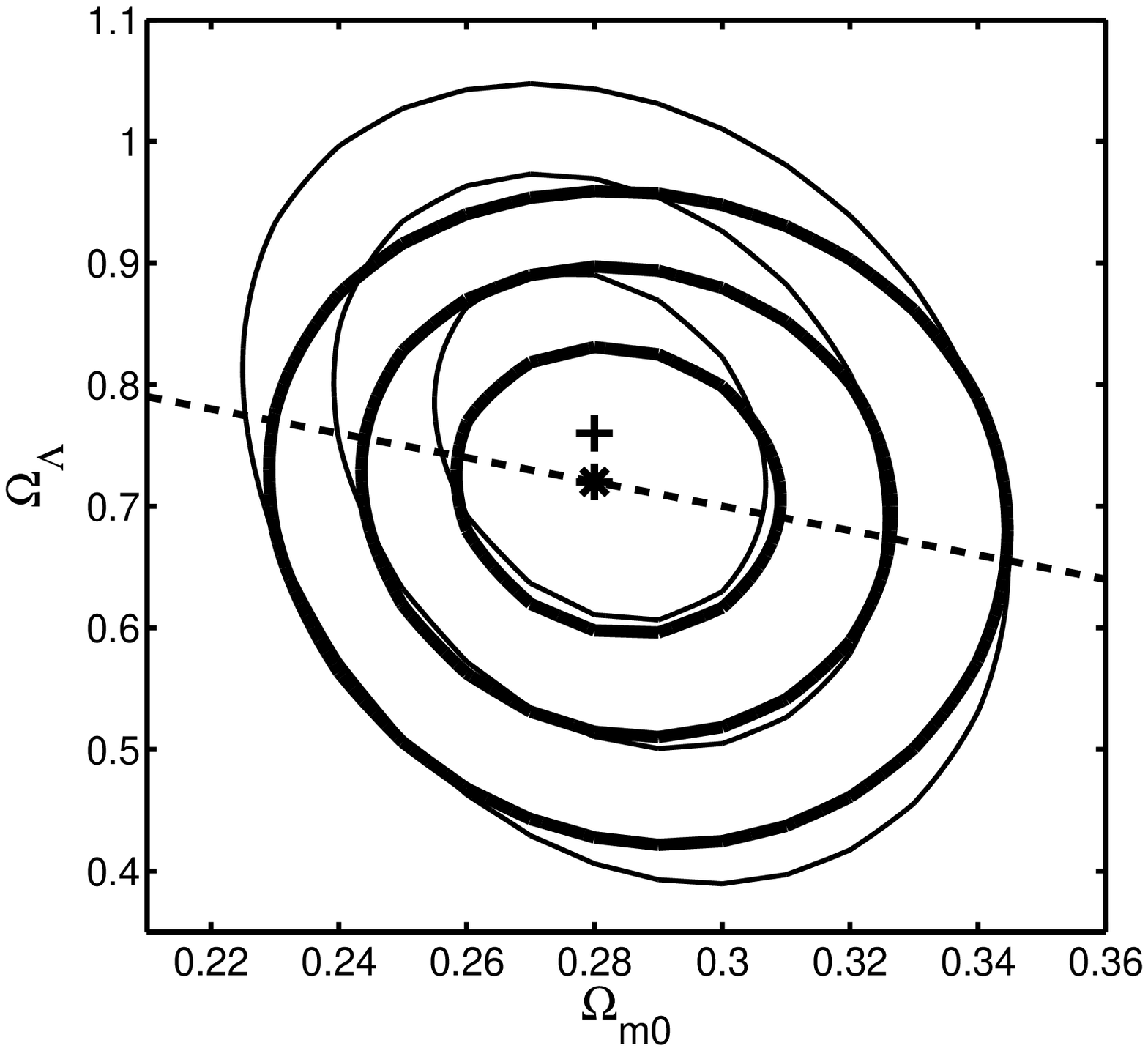}

 \caption{
Thick (thin) solid lines are 1, 2, and 3 $\sigma$ constraint
contours for the $\Lambda$CDM model from a joint analysis of the BAO
and SNeIa (with systematic errors) data, with (and without) the
$H(z)$ data. The cross (``+'') marks the best-fit point determined
from the joint sample without the $H(z)$ data at $\Omega_{m0}=0.28$
and $\Omega_{\Lambda}=0.76$ with $\chi^2_{\rm min}=531$. The star
(``$*$'') marks the best-fit point determined from the joint sample
with the $H(z)$ data at $\Omega_{m0}=0.28$ and
$\Omega_{\Lambda}=0.72$ with $\chi^2_{\rm min}=541$. The dashed
sloping line corresponds to spatially-flat models.}
\label{fig:LCDM_com}
\end{figure}

\begin{figure}[t]
\centering
  \includegraphics[angle=0,width=90mm]{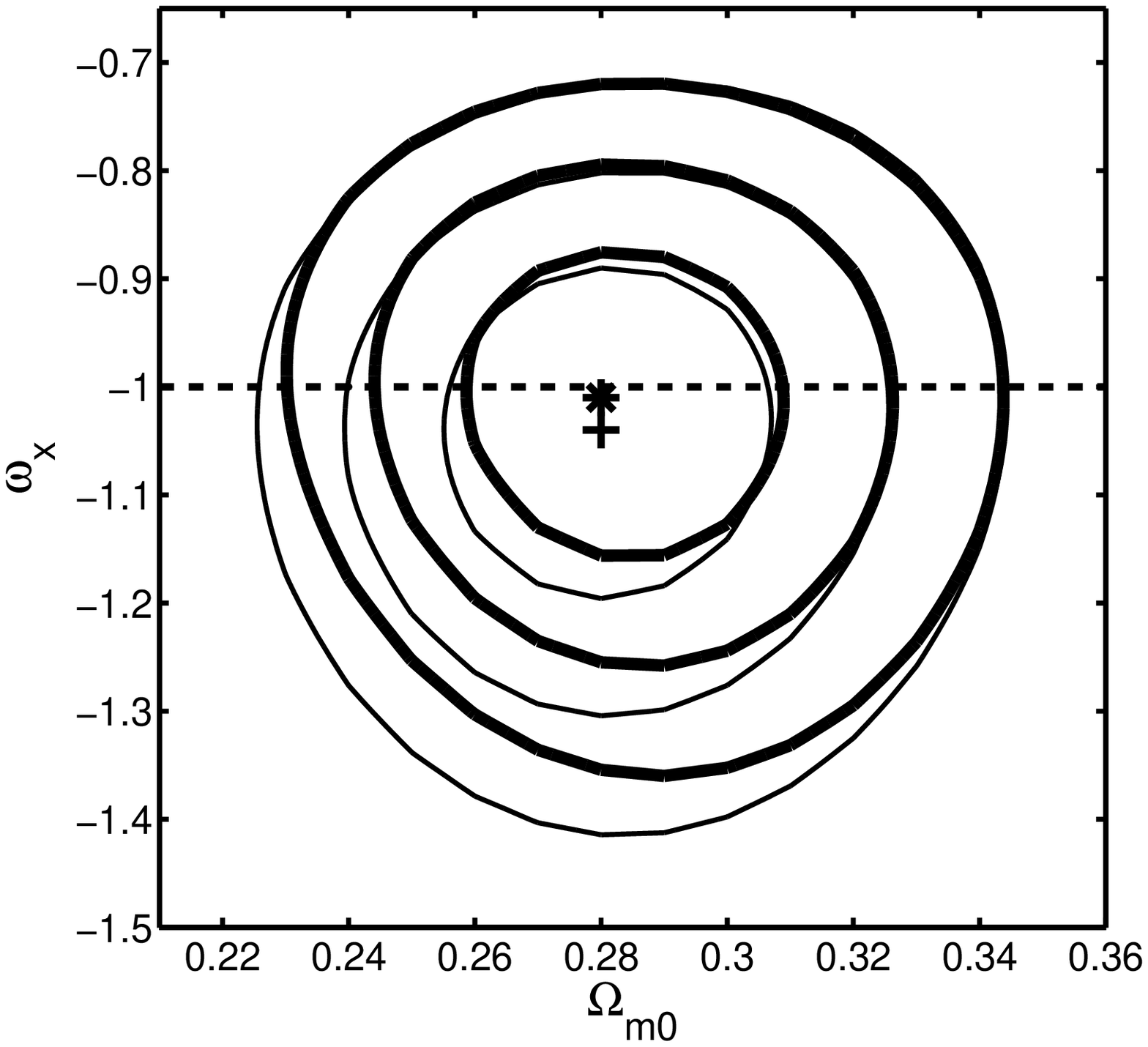}

 \caption{
Thick (thin) solid lines are 1, 2, and 3 $\sigma$ constraint
contours for the XCDM parametrization from a joint analysis of the
BAO and SNeIa (with systematic errors) data, with (and without) the
$H(z)$ data. The cross (``+'') marks the best-fit point determined
from the joint sample without the $H(z)$ data at $\Omega_{m0}=0.28$
and $\omega_X=-1.04$ with $\chi^2_{\rm min}=531$. The star (``$*$'')
marks the best-fit point determined from the joint sample with the
$H(z)$ data at $\Omega_{m0}=0.28$ and $\omega_X=-1.01$ with
$\chi^2_{\rm min}=541$. The dashed horizontal line at $\omega_X =
-1$ corresponds to spatially-flat $\Lambda$CDM models.}
\label{fig:XCDM_com}
\end{figure}

\begin{figure}[t]
\centering
  \includegraphics[angle=0,width=90mm]{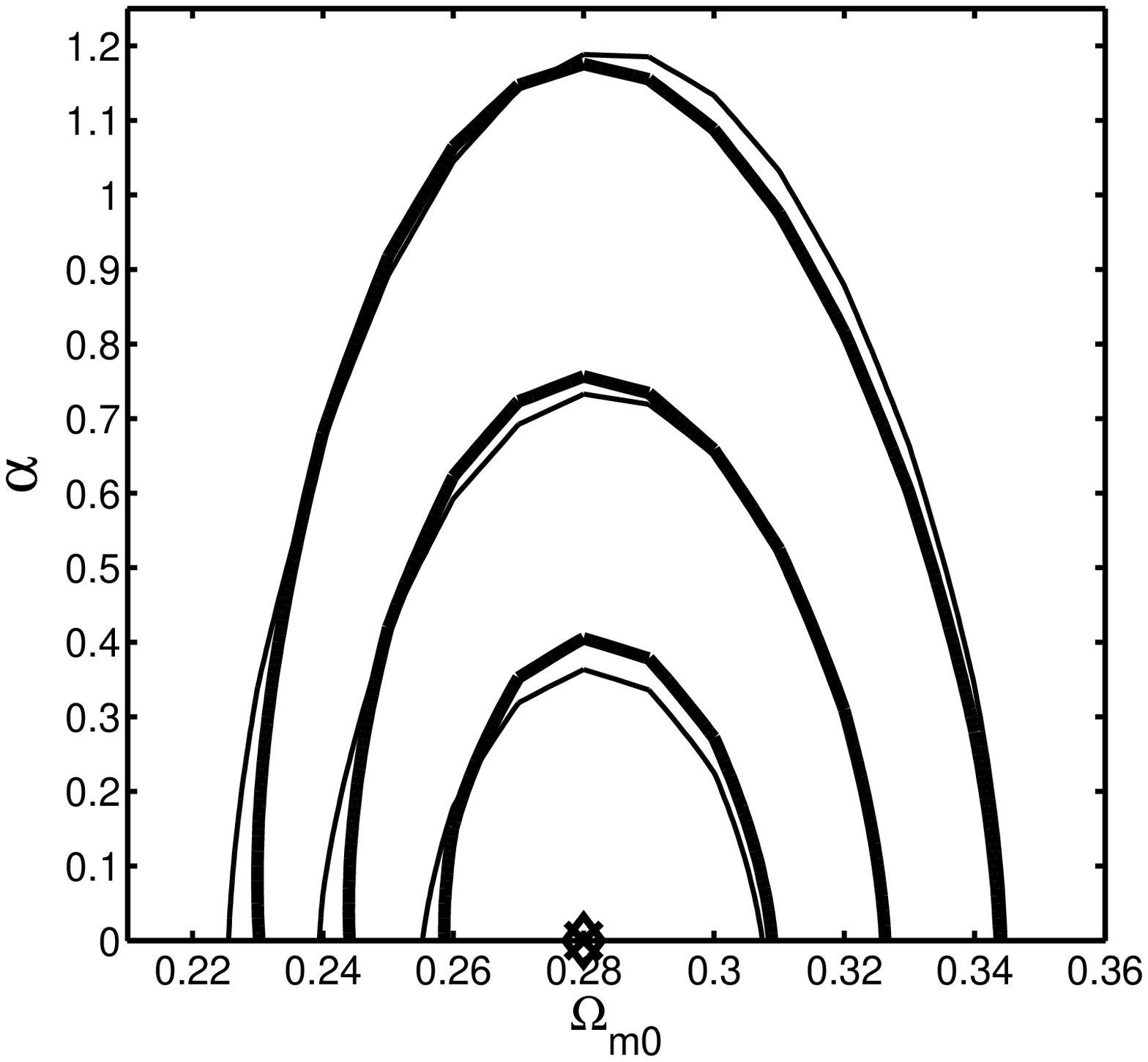}

 \caption{
Thick (thin) solid lines are 1, 2, and 3 $\sigma$ constraint
contours for the $\phi$CDM model from a joint analysis of the BAO
and SNeIa (with systematic errors) data, with (and without) the
$H(z)$ data. The cross (``$\times$'') marks the best-fit point
determined from the joint sample without the $H(z)$ data at
$\Omega_{m0}=0.28$ and $\alpha=0$ with $\chi^2_{\rm min}=531$. The
diamond (``$\diamondsuit$'') marks the best-fit point determined
from the joint sample with the $H(z)$ data at $\Omega_{m0}=0.28$ and
$\alpha=0.0$ with $\chi^2_{\rm min}=541$. The $\alpha = 0$
horizontal axis corresponds to spatially-flat $\Lambda$CDM models.}
\label{fig:phiCDM_com}
\end{figure}

Figures \ref{fig:LCDM_com}--\ref{fig:phiCDM_com} show the
constraints on the cosmological parameters for the $\Lambda$CDM and
$\phi$CDM models and the XCDM parametrization from a joint analysis
of the BAO and SNeIa data, as well as from a joint analysis of the
BAO, SNeIa and $H(z)$ data. Adding the $H(z)$ data tightens up the
constraints somewhat, most significantly in the $\Lambda$CDM case
(Fig.\ 4) and least so for the $\phi$CDM model (Fig.\ 6).

\begin{figure}[t]
\centering $\begin{array}{cc}
\includegraphics[width=0.4\textwidth]{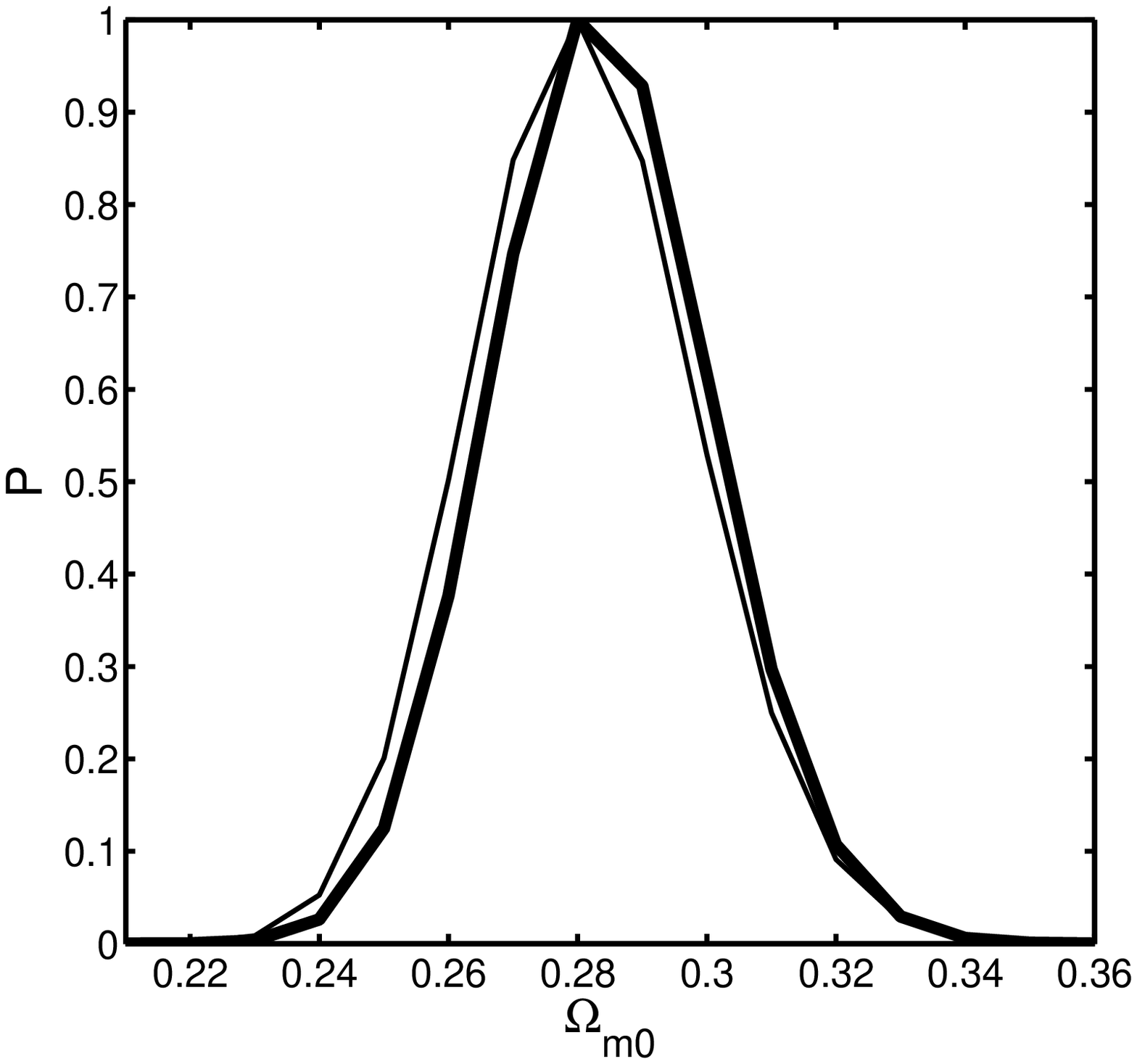}\\
\includegraphics[width=0.4\textwidth]{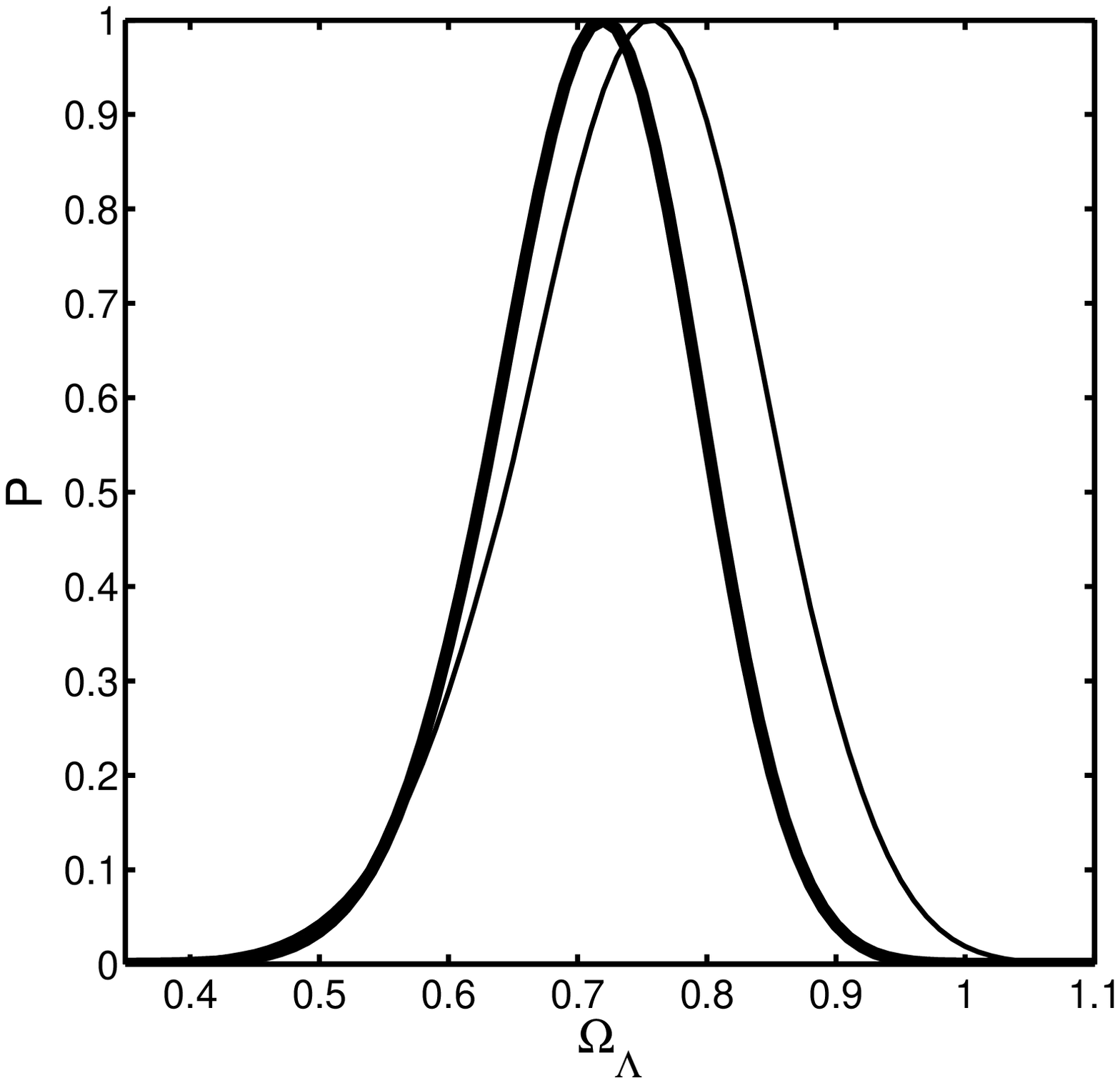}
\end{array}$
 \caption{One-dimensional marginalized distribution probabilities
of the cosmological parameters for the LCDM model. Thick (thin)
lines are results from a joint analysis of the BAO and SNeIa (with
systematic errors) data, with (and without) the $H(z)$ data.}
\label{fig:Pro_LCDM}
\end{figure}

\begin{figure}[t]
\centering $\begin{array}{cc}
\includegraphics[width=0.4\textwidth]{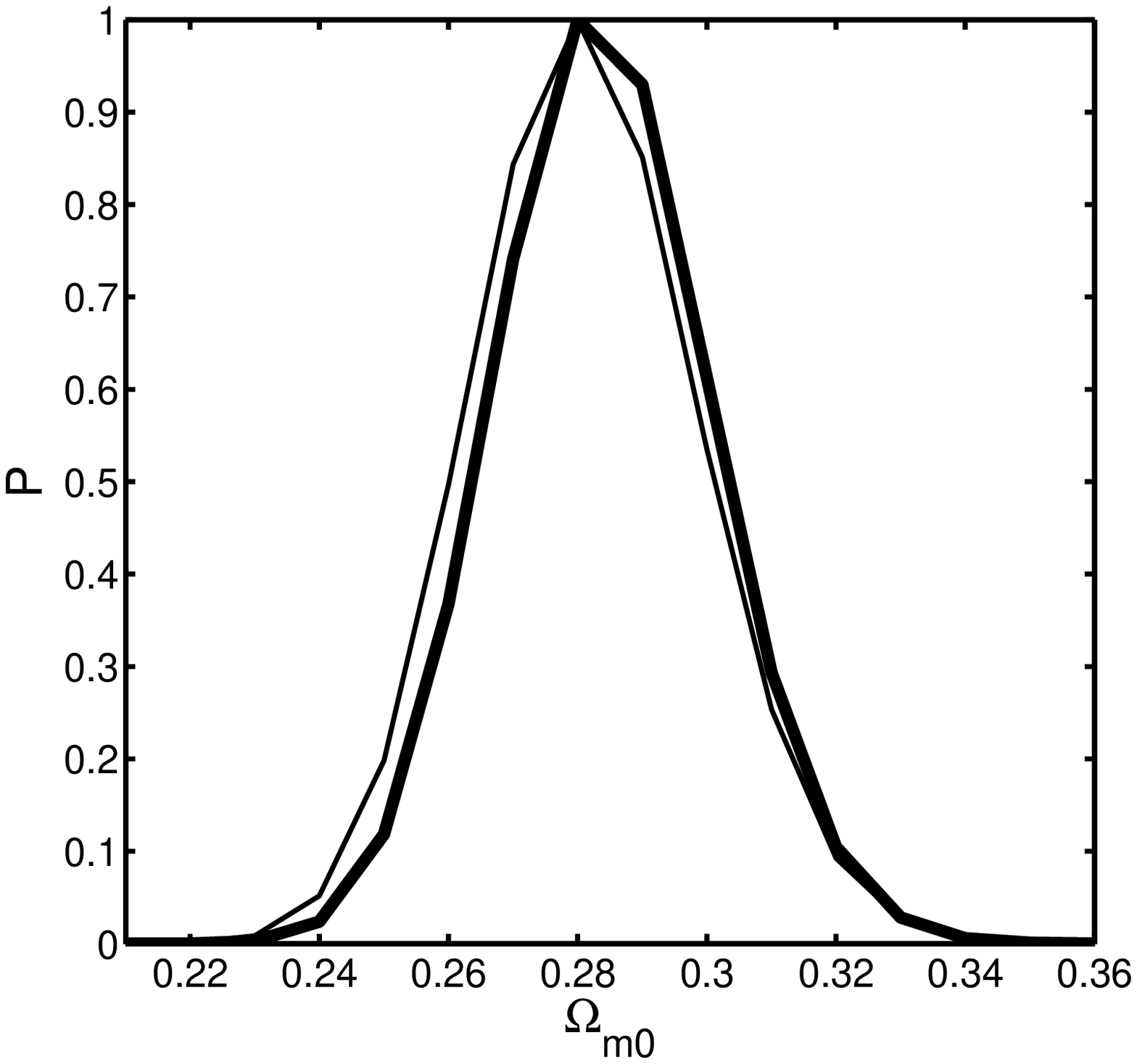}\\
\includegraphics[width=0.4\textwidth]{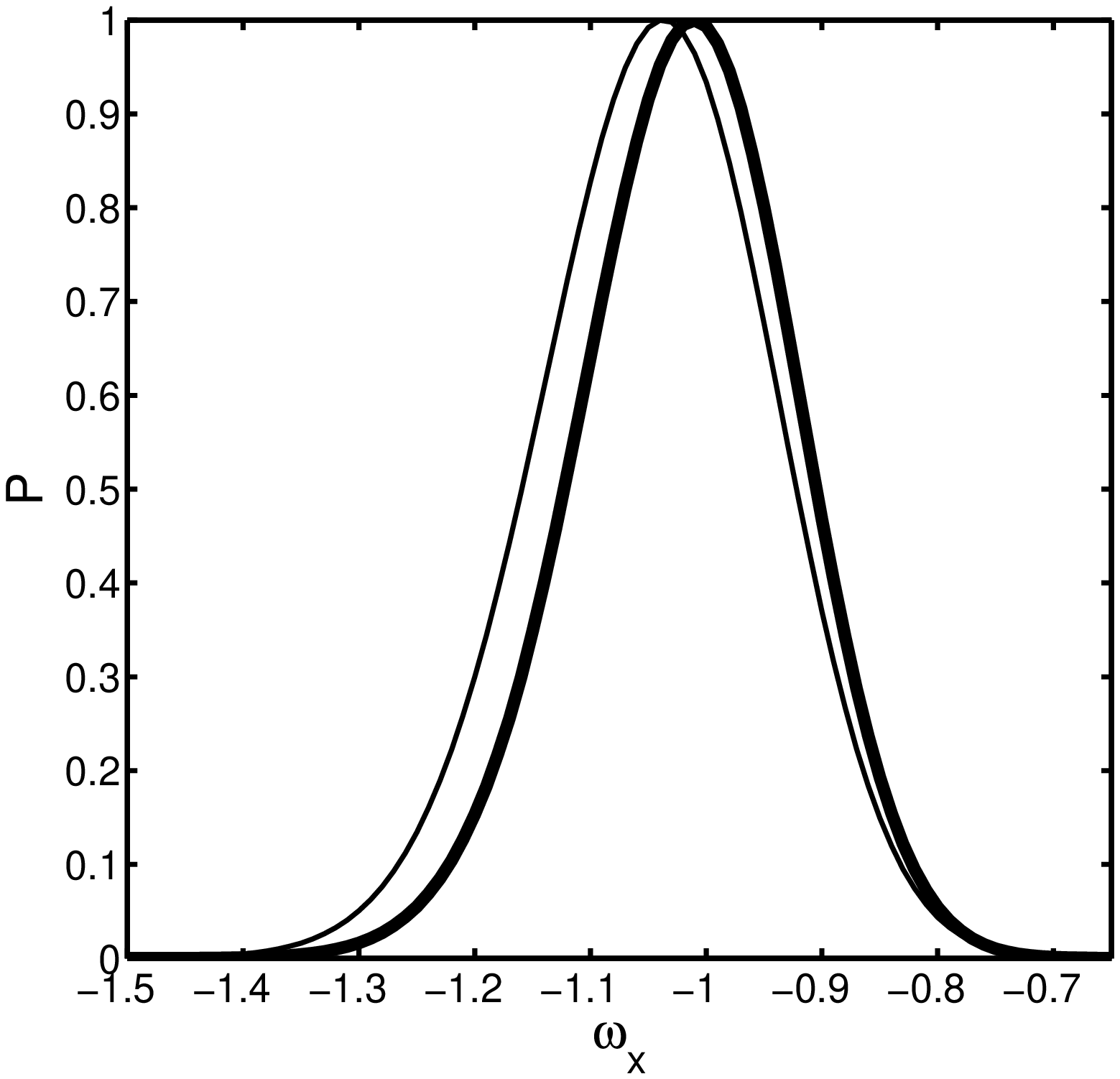}
\end{array}$
 \caption{One-dimensional marginalized distribution probabilities
of the cosmological parameters for the XCDM parametrization. Thick
(thin) lines are results from a joint analysis of the BAO and SNeIa
(with systematic errors) data, with (and without) the $H(z)$ data.}
\label{fig:Pro_XCDM}
\end{figure}

\begin{figure}[t]
\centering $\begin{array}{cc}
\includegraphics[width=0.4\textwidth]{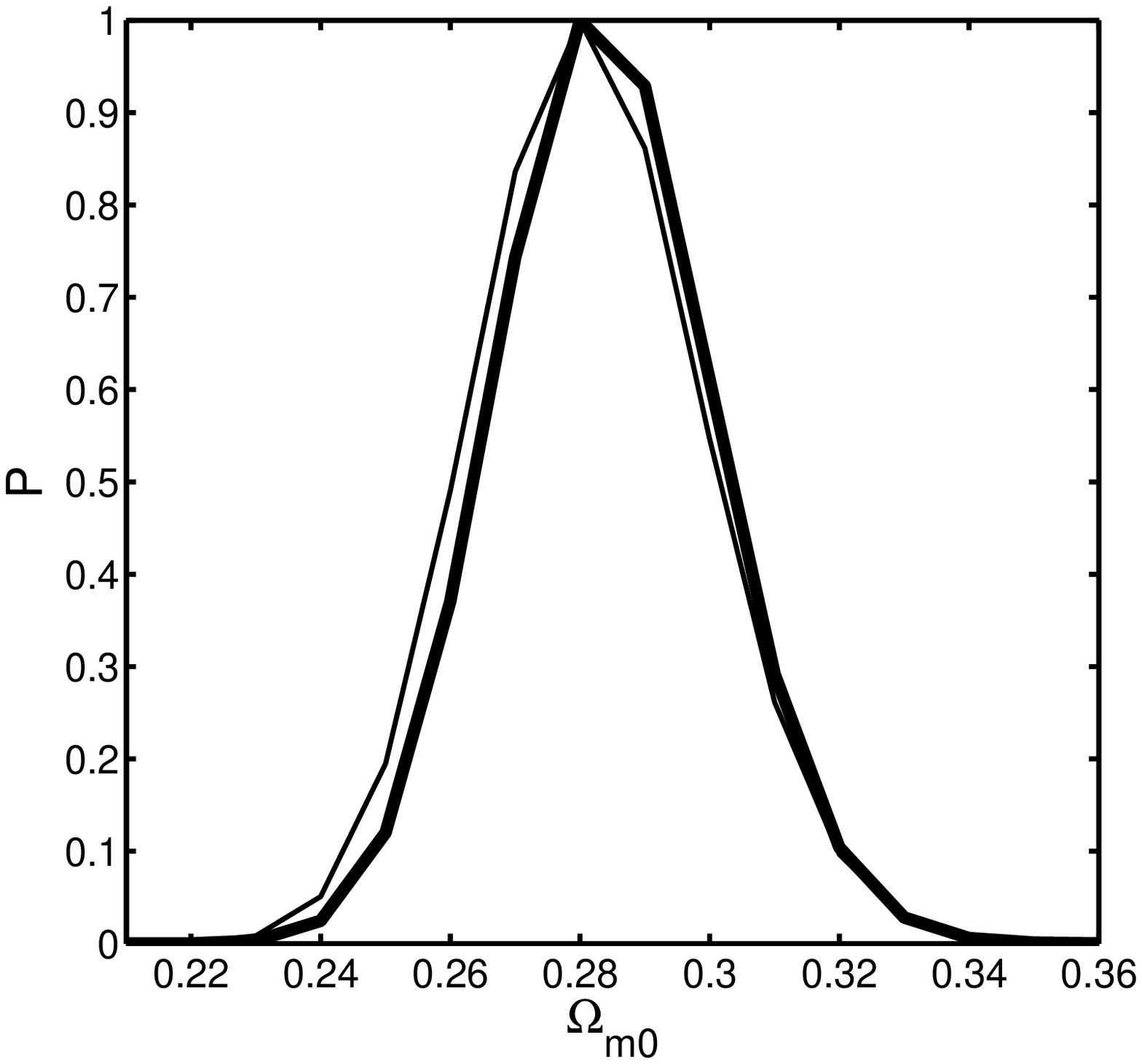}\\
\includegraphics[width=0.4\textwidth]{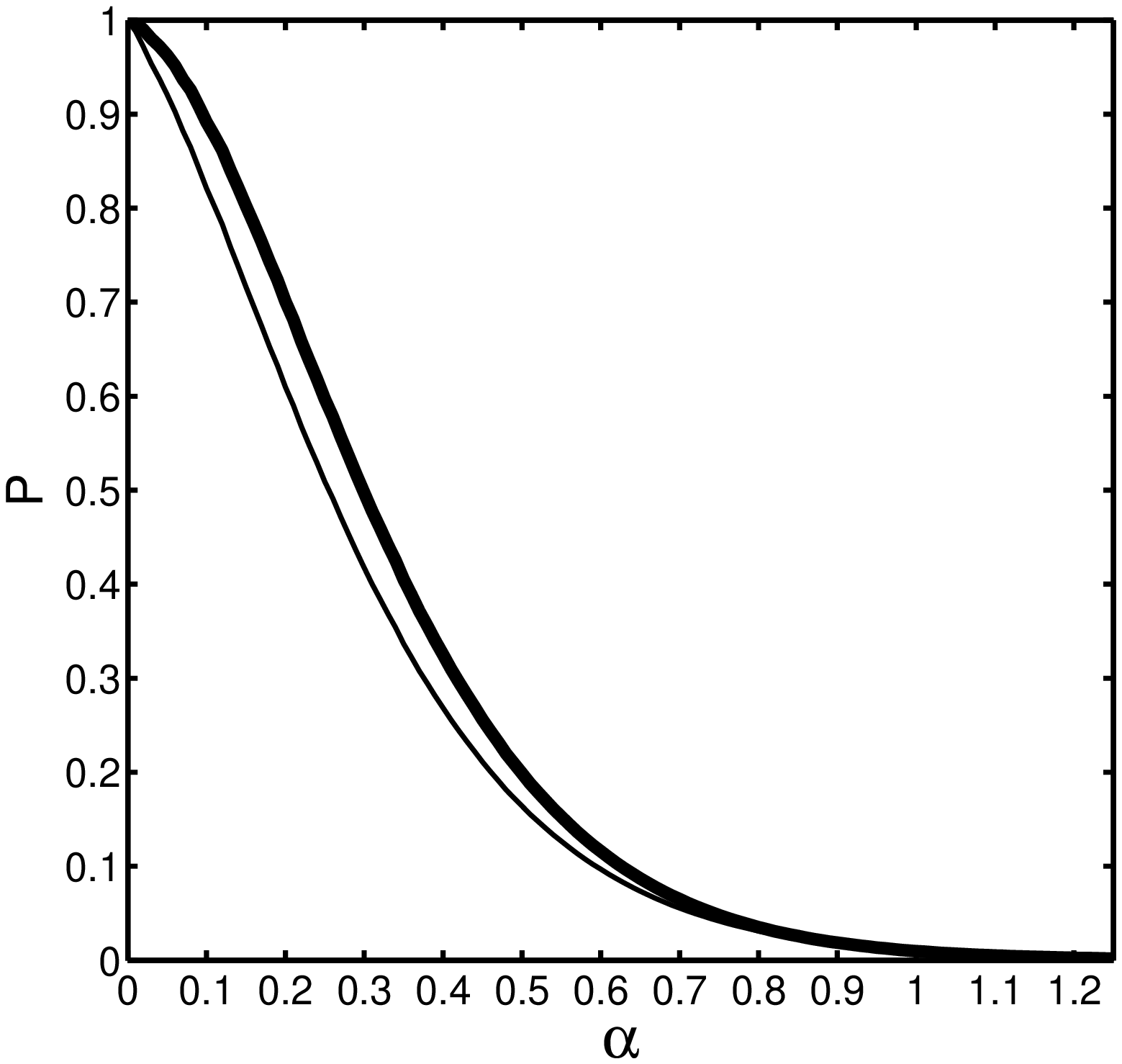}
\end{array}$
 \caption{One-dimensional marginalized distribution probabilities
of the cosmological parameters for the $\phi$CDM  model. Thick
(thin) lines are results from a joint analysis of the BAO and SNeIa
(with systematic errors) data, with (and without) the $H(z)$ data.}
\label{fig:Pro_phiCDM}
\end{figure}

Figures \ref{fig:Pro_LCDM}--\ref{fig:Pro_phiCDM} display the
one-dimensional marginalized distribution probabilities of the
cosmological parameters for the three cosmological models considered
in this work, derived from a joint analysis of the BAO and SNeIa
data, as well as from a joint analysis of the BAO, SNeIa and $H(z)$
data. The marginalized 2 $\sigma$ intervals of the cosmological
parameters are presented in Table \ref{tab:intervals}.

The combination of BAO and SNeIa data gives tight constraints on the
cosmological parameters. Adding the currently-available $H(z)$ data
to the mix does shift the constraint contours, however the effect is
significant only in some parts of parameter space for only some of
the models we study. While useful, current $H(z)$ data do not have
enough weight to significantly affect the combined BAO and SNeIa
results in most parts of model-parameter space. The $H(z)$ data have
a little more weight than currently-available gamma-ray burst
luminosity measurements (\cite{Samushia&Ratra2010}, Figs.\ 4--6 and
10--12).

In summary, the $H(z)$ data considered here are very consistent with
the predictions of a spatially-flat cosmological model with energy
budget dominated by a time-independent cosmological constant.
However, the data do not rule out time-evolving dark energy,
although they do require that it not vary rapidly.

\section{Conclusion}
\label{summary}

We have shown that the Hubble parameter versus redshift data from
S10 and G09 can give interestingly restrictive constraints on
cosmological parameters. The resulting constraints are compatible
with those derived from other current data, thus strengthening
support for the current ``standard'' cosmological model. The $H(z)$
data constraints are approximately as restrictive as those that
follow from currently-available gamma-ray burst luminosity data and
lookback time observations, and more restrictive than those from
currently-available galaxy cluster angular size data. They are,
however, much less restrictive than those that follow from a
combined analysis of BAO peak length scale and SNeIa apparent
magnitude data.

The spatially-flat $\Lambda$CDM model, currently dominated by a
constant cosmological constant, provides a good fit to the data we
have studied here. However, these data do not rule out a
time-evolving dark energy.

As discussed in \cite{Ma2011}, future high-$z$, high-accuracy $H(z)$
determinations from BAO observations will likely result in
cosmological parameter constraints comparable to those that follow
from SNeIa data.

\acknowledgments
Yun Chen thanks Data Mania for useful discussions. YC was supported
by the China State Scholarship Fund No.\ 2010604111 and the Ministry
of Science and Technology national basic science program (project
973) under grant No.\ 2007CB815401. BR was supported by DOE grant
DEFG03-99EP41093.

\clearpage
\begin{table}
\begin{center}
\begin{tabular}{lcl}
\hline\hline
$z$ & $H(z)$ & $\sigma_{H}$ \\
\tableline
0.1 &69 & 12 \\
0.17& 83 & 8\\
 0.24& 79.69 & 2.65\\
0.27& 77& 14\\
 0.4& 95& 17\\
0.43& 86.45& 3.68\\
0.48&97 &60\\
 0.88& 90& 40\\
 0.9& 117 & 23\\
1.3& 168& 17\\
 1.43& 177& 18\\
 1.53& 140& 14\\
 1.75& 202& 40\\
\hline\hline
\end{tabular}
\end{center}
\caption{Hubble parameter versus redshift data from S10 and G09.
Where $H(z)$ and $\sigma_{H}$ are in km s$^{-1}$
Mpc$^{-1}$.}\label{tab:Hz}
\end{table}

\begin{table}
\vspace*{12pt}
\begin{center}
\begin{tabular}{lcc}
\hline\hline
Model & BAO + SNeIa  & $H(z)$ + BAO + SNeIa\\
\tableline

$\Lambda$CDM & $0.24<\Omega_{m0}<0.33$
& $0.24<\Omega_{m0}<0.33$  \\
& $0.5<\Omega_{\Lambda}<0.97$ & $0.51<\Omega_{\Lambda}<0.9$\\
\tableline

 XCDM & $0.24<\Omega_{m0}<0.33$
& $0.24<\Omega_{m0}<0.33$\\
&$-1.30<\omega_X<-0.80$ & $-1.26<\omega_X<-0.79$\\
\tableline

$\phi$CDM & $0.24<\Omega_{m0}<0.33$
& $0.24<\Omega_{m0}<0.33$ \\
& $0<\alpha<0.73$ & $0<\alpha<0.76$ \\
\hline\hline
\end{tabular}
\end{center}
\caption{Two standard deviation bounds on cosmological
parameters.}\label{tab:intervals}
\end{table}

\end{document}